\documentclass[10pt,twocolumn,letterpaper]{article}

\usepackage[pagenumbers]{cvpr} 

\usepackage{silence}
\usepackage{graphicx}
\usepackage{amsmath}
\usepackage{amssymb}
\usepackage{booktabs}
\usepackage{dblfloatfix}
\usepackage{url}
\usepackage{float}
\usepackage[accsupp]{axessibility}

\usepackage[pagebackref,breaklinks,colorlinks]{hyperref}

\usepackage[capitalize]{cleveref}
\crefname{section}{Sec.}{Secs.}
\Crefname{section}{Section}{Sections}
\Crefname{table}{Table}{Tables}
\crefname{table}{Tab.}{Tabs.}

\def\cvprPaperID{10994} 
\def\confName{CVPR}
\def\confYear{2022}

\newcommand{\myparagraph}[1]{\smallskip\noindent\textbf{#1}}
\begin{document}

\title{Temporal Context Matters: Enhancing Single Image Prediction with Disease Progression Representations}

\author{\large
Aishik Konwer\textsuperscript{1},\hskip 1em
Xuan Xu\textsuperscript{1},\hskip 1em
Joseph Bae\textsuperscript{2}, \hskip 1em 
Chao Chen\textsuperscript{2},\hskip 1em
Prateek Prasanna\textsuperscript{2}
\\
\textsuperscript{1}Department of Computer Science, Stony Brook University \hskip 1em
\\
\textsuperscript{2}Department of Biomedical Informatics, Stony Brook University \hskip 1em
\\
{\tt\small \{akonwer, xuaxu\}@cs.stonybrook.edu \{joseph.bae, Chao.Chen.1, Prateek.Prasanna\}@stonybrook.edu}  
}
\maketitle

\begin{abstract}
Clinical outcome or severity prediction from medical images has largely focused on learning representations from single-timepoint or snapshot scans. It has been shown that disease progression can be better characterized by temporal imaging. We therefore hypothesized that outcome predictions can be improved by utilizing the disease progression information from sequential images. We present a deep learning approach that leverages temporal progression information to improve clinical outcome predictions from single-timepoint images. In our method, a self-attention based Temporal Convolutional Network (TCN) is used to learn a representation that is most reflective of the disease trajectory. Meanwhile, a Vision Transformer is pretrained in a self-supervised fashion to extract features from single-timepoint images. The key contribution is to design a recalibration module that employs maximum mean discrepancy loss (MMD) to align distributions of the above two contextual representations. We train our system to predict clinical outcomes and severity grades from single-timepoint images. Experiments on chest and osteoarthritis radiography datasets demonstrate that our approach outperforms other state-of-the-art techniques. 
\end{abstract}

\section{Introduction}
\label{sec:intro}

Predicting clinical outcomes from medical images is a long standing goal in the medical vision community \cite{outcome1, outcome2, outcome3, outcome4}. For the past half a decade, researchers have employed various deep neural networks (DNNs) \cite{dnn1, dnn2, dnn3} to improve diagnostic and prognostic performance. Previously, DNNs were trained from scratch \cite{artcnn2} for classification and detection tasks on various medical imaging datasets. These multi-organ datasets can range from 2D radiographs (x-rays) \cite{rad1} to 3D magnetic resonance imaging (MRI) \cite{mri1} or computerized tomography (CT) \cite{ct1} scans. 
More recent frameworks have employed knowledge distillation \cite{tl1, kd1} and self-supervision techniques \cite{tl2, tl3} 
to pretrain models which are then finetuned on limited  medical imaging data. This has led to improved model performance. 

However, most medical imaging datasets contain only single-timepoint or `snapshot' images. Although a snapshot image plays an essential role for describing a disease, \textit{sequential scans provide a more comprehensive characterization of the evolution and prognosis of a pathology}. 
The temporal evolution of imaging biomarkers are highly correlated with disease progression trajectory. We hypothesize that this rich underlying domain information can be leveraged by deep learning approaches to make accurate predictions about the disease trajectory even when temporal data is limited/unavailable. 

In practice, temporal medical data can be very limited because patients are often lost to follow-up or suffer from chronic diseases with infrequent re-evaluations of their condition. Temporal models usually overfit on these small datasets leading to poor generalizability. Hence, they are limited in their use as a standalone source for training recurrent neural networks (RNN), Temporal ConvNets (TCN), etc. Recently there have been many deep learning--based works that aim to learn representations from sequential medical imaging data \cite{I1, I2, I3, I4}. The bottleneck of limited training samples is evident in all of them. Unlike the video vision community where the presence of large scale temporal datasets facilitates temporal modeling approaches, pursuing similar problems (for e.g., future timepoint severity prediction, object evolution) in medical imaging scenario is technically challenging. 

In this work we propose to learn disease progression patterns from limited temporal imaging data, and use this auxiliary knowledge to enhance predictive performance of methods that use snapshot scans. Since the representations are obtained from two different domains - snapshot and temporal - the challenges lie in how to optimally adapt and align these feature distributions. Because each image in a temporal sequence contributes unequally, we first extract an `optimal' embedding of the entire sequence. An `optimal' embedding should retain maximum information focused on the key transition stages over the course of a disease. The temporal feature representation can then be aligned against a snapshot feature. Our next step involves employing an appropriate feature matching technique to re-calibrate the two different domain representations (snapshot and temporal). We build a framework that leverages partially available temporal data to re-calibrate the representations learned by the single-timepoint pipeline. A temporal network that employs multi-head self-attention at each layer is incorporated in our architecture. We eventually obtain a global attention distribution that aids in selecting an optimal representation from the whole sequence. Meanwhile, a vision transformer is pretrained in a self-supervised fashion to extract features from snapshot images. Finally, during the finetuning phase, maximum mean discrepancy (MMD) loss is proposed as a feature matching tool to minimize the distance between the two representations. 

The main contributions of this work are as follows:
\begin{itemize}
    \item This is the first work that learns representations from limited temporal medical images, and eventually utilizes them to improve clinical prediction tasks from single-timepoint datasets.
    \item We use a Temporal ConvNet that employs hierarchical attention to obtain the most optimal representation of a temporal image sequence, so that it can be compared with the features from a single image--based pipeline.
    \item In our study, intermediate representations are available from temporal and snapshot images. We propose to use MMD loss for the first time in this domain, to align the snapshot feature space with the optimal temporal representations selected through an attention mechanism.
\end{itemize}


\section{Related works}
\label{sec:related}

\begin{figure}[t]
  \centering
  \includegraphics[width=1\linewidth]{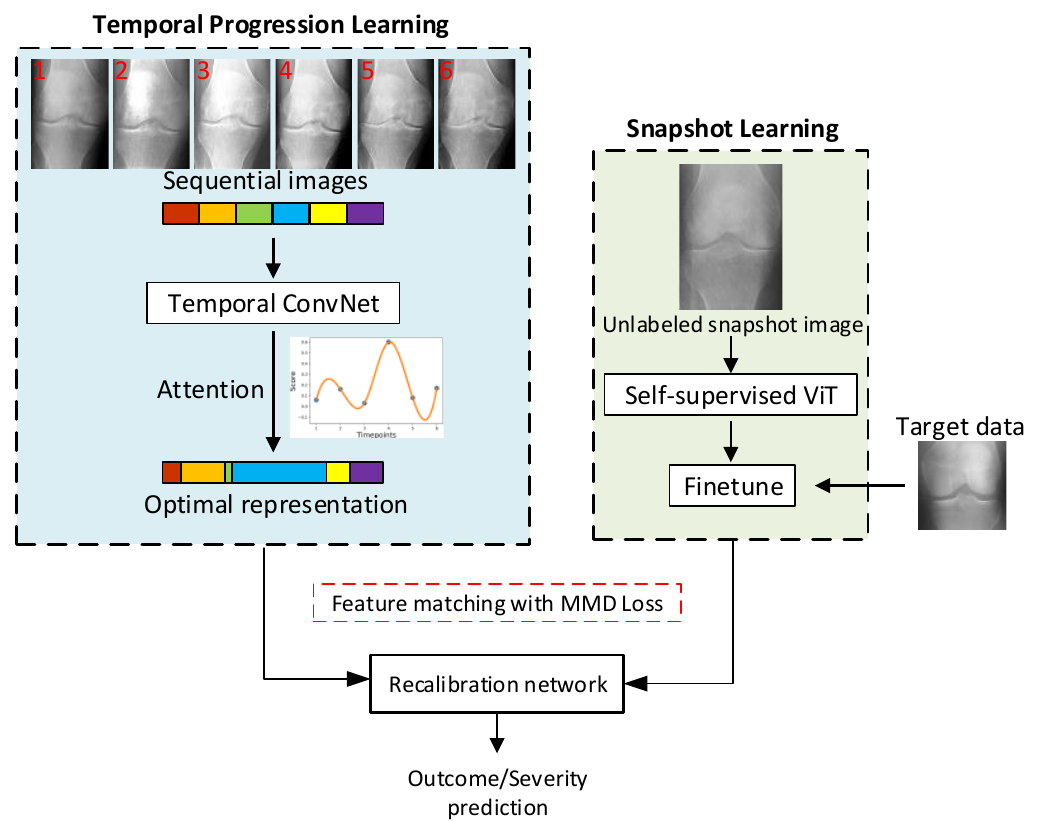}
   \caption{\textbf{Overview of our proposed method.} Temporal learning module learns the optimal representation from sequential images. Snapshot learning extracts representations from snapshot images. The Recalibration network aligns the two contextual representations using MMD loss.}
  \label{overview}
\end{figure}

\subsection{Temporal modeling of disease progression}
In the current era of precision medicine, temporal modeling of disease progression is an important field of research. To learn disease trajectory from sequential imaging data, medical vision researchers have typically applied various recurrent neural networks (RNNs) widely used by the video analytics community. Recurrent convolutional models \cite {T1, lrcn, lstm} were the first to process variable sequence inputs and exploit their long-term dynamics across timepoints. Along this direction, many recent techniques have been developed for recognizing human action in videos \cite{TDN,TPN}. Temporal Difference Network \cite{TDN} simultaneously learns both short and long-term motion information to improve action recognition performance. Yang et. al \cite{TPN} introduced a pyramid network (TPN) that utilized both slow and fast tempos through a single multi-level architecture by unifying features from different hierarchies. GESTURES \cite{GESTURES} pretrained a spatio-temporal CNN on action recognition data to extract representations from seizure videos. Konwer et. al \cite{I4} used an architecture inspired by CorrRNN \cite{cornn} for predicting disease severity at a future timepoint. To do so, a correlation module was integrated within a Gated Recurrent Unit (GRU) to exploit the disease correlation among different zonal patches. Zhang et. al \cite{convlstm} captured both the 3D spatial context and temporal dynamics of a growing tumor by employing a spatio-temporal ConvLSTM framework. 

Temporal ConvNets (TCN) have been shown to be more effective over LSTM-based methods in tasks like action segmentation and recognition \cite{TCN1,TCN2}, but are yet to be fully explored for medical imaging. Recently, TCN showed significant improvement over recurrent models in Alzheimer's disease detection \cite{I1}. In our work, we make use of a TCN-based architecture to select the optimal representation of the entire temporal sequence. This requires incorporating attention within the temporal approach. A TCN-based architecture is more explainable than RNNs because attention can be incorporated at various levels - kernel, layer-wise or globally \cite{AttTCN,AttTCN2}. This motivates us to use a hierarchical attention--based TCN architecture for the temporal section of our framework. Though transformers have recently shown improved performance over TCN, limited datasets restrict us to a lightweight transformer alternative like TCN.

\subsection{Feature and probability distribution level losses}
Distance measures that match two image distributions, are generally found to operate at two levels - 1) posterior probability and 2) feature space. KL-divergence loss \cite{kloriginal} and Bhattacharya distance \cite{bhattoriginal} are two popular measures in the former category that have been used for vision problems like visual tracking and image segmentation \cite{kl1,kl2,bhatt1,bhatt2}. However, in our case, the image representations to be compared are derived from two different contexts, i.e, temporal and snapshot. Also, the tasks performed to obtain them are not similar. It is a hard imposition to match the underlying probability distributions. Hence, we used the second category losses - MMD and CORAL loss to align the image feature distributions. \cite{D5,D6,D8,D9}. These are popular metrics for feature adaptation between two types of data. MMD is widely used to interpret tasks like neural style transfer \cite{style}, or improving unsupervised image generation \cite{mmdgan}. These losses minimize the distance between various statistical measures (first and second moments) of image embeddings.   

\subsection{Self-supervised Transformers in medical imaging}
The introduction of Vision transformers (ViT)~\cite{Transformer1} proved that even without using CNNs, compelling classification performance could be achieved in supervised approaches solely from a sequence of image patches. ViT exploits the long-range spatial dependencies in images, previously underutilized by CNN models. After pretraining, ViT can be finetuned on downstream medical datasets to achieve improved results in classification and segmentation tasks~\cite{V1,V2,transmed,transbts}. This is perfectly suited for analyzing medical images, where intra-image disease manifestations can be characterized more comprehensively using ViTs. Hence, we propose to use a ViT as a feature extractor for snapshot images in our framework. A major disadvantage of ViT is that their pretraining necessitates large scale datasets. 

Self-supervised learning approaches \cite{SS1,SS2} have made significant advances in recent years, improving the ability to learn image representations even from smaller datasets. This is achieved by training models on well-designed proxy goals that do not require manual annotations. Examples of these goals include discriminative tasks such as predicting image rotation \cite{SS3}, solving multimodal jigsaw puzzles \cite{SS4}, etc. In this work, we focus on training ViT in a self-supervised paradigm by leveraging large-scale, albeit unannotated, snapshot medical images. 





\begin{figure*}[t]
  \centering
  \includegraphics[width=0.7\linewidth]{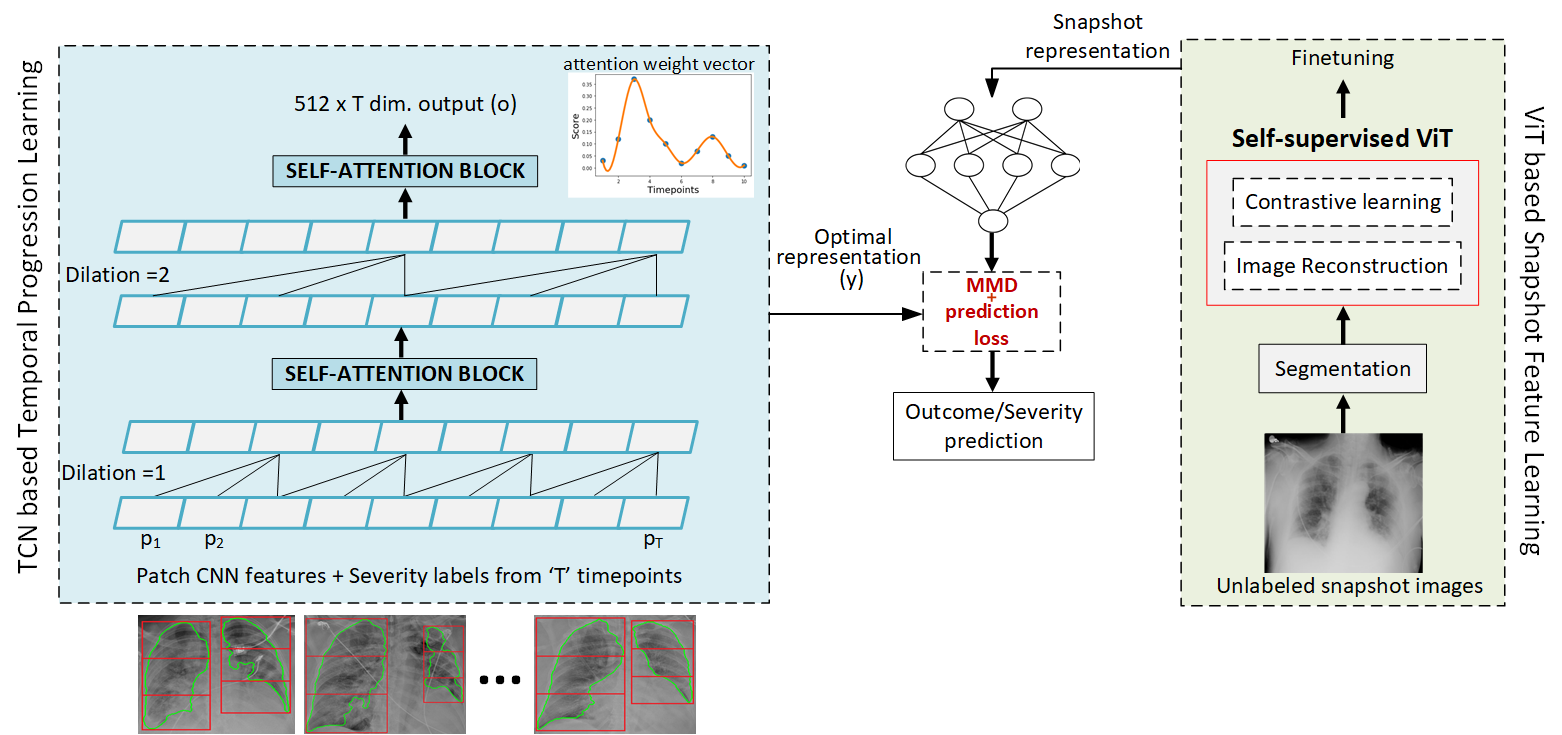}
    \caption{\textbf{Illustration of the proposed architecture}. Self-attention block after each layer of TCN, helps to obtain the optimal representation from each sequence of temporal images. Meanwhile, snapshot image representations are generated through a self-supervised ViT. The snapshot and temporal representations are aligned using MMD loss while training a downstream neural network}
    \label{architecture}
\end{figure*}

\section{Methodology}
\label{sec:method}

Given a snapshot image of an individual patient, we predict their clinical outcomes/disease severity by utilizing the disease progression information previously learnt from a small temporal dataset with sequential scans. An overview of our proposed framework is shown in \cref{overview}. 
The key idea is to use the feature representations of the temporal images to recalibrate the representation of snapshot images. In the training stage, besides finetuning the snapshot representations through the standard loss (e.g., cross-entropy loss or ordinal loss, depending on the task), we also match them with the temporal representations at a distribution level, using the MMD loss. At the inference time, the recalibrated snapshot representations are used for the final prediction. This recalibration strategy, as well as the training losses, are elaborated in section \ref{ssec:method2}.
Prior to this, we use state-of-the-art methods for obtaining both the snapshot and the temporal representations. First, we extract hidden representations from temporal images using a hierarchical TCN. This step selects the best representation from each sequence. Detailed description is provided in section \ref{ssec:method1}. As for the snapshot representations, we use a pretrained self-supervised vision transformer (see section \ref{ssec:method3}). 

\subsection{Representation Recalibration network and training losses}
\label{ssec:method2}
We first introduce the key component of our method -- the \emph{recalibration network}. It uses representations of temporal images to recalibrate representations of snapshots. More details of how these representations are obtained will be explained in later sections. Here we assume snapshot/single-timepoint image embeddings $x^{s}\in\mathbb{R}^{512}$. They are fed as input to the \emph{recalibration network}, i.e., a multilayer perceptron (MLP) with 1 hidden layer. We also have temporal representations $y^{t}\in\mathbb{R}^{512}$ from temporal sequences. The temporal representations are only used in the training stage to recalibrate snapshot representations. To train the recalibration network, we use both the standard prediction loss, $\mathcal{L}_{prediction}$, and the MMD loss, $\mathcal{L}_{MMD}$, to match the snapshot representations to the temporal representations.  
\begin{equation}
  \mathcal{L} = \lambda_{1}\mathcal{L}_{MMD} + \mathcal{L}_{prediction}
  \label{eq:final-all}
\end{equation}

Given two distributions $P_s$ and $P_t$ for snapshot and temporal data, by mapping the data into a reproducing kernel Hilbert space (RKHS) using function $\phi(\cdot)$, the MMD between the two distributions is calculated as  

\begin{equation*}
 \small
{{\rm{MMD}}^2}(s,t) = \mathop {sup }\limits_{{{\left\| \phi  \right\|}_{\cal H}} \le 1} \left\| {{E_{{x^s} \sim P_s}}\left[ {\phi ({x^s})} \right] - {E_{{y^t} \sim P_t}}\left[ {\phi ({y}^t)} \right]} \right\|_{\cal H}^2
  \label{eq:mmd}
\end{equation*}
where ${E_{{x^s} \sim P_s}}\left[  \cdot  \right]$ denotes the expectation with regard to the distribution~$P_s$, and ${\left\| \phi  \right\|_{\cal H}} \le 1$ defines a set of functions in the unit ball of a RKHS, ${\cal H}$. In our setting, the problem is simplified as we assume the data are all represented in the same latent space with Euclidean metric. The MMD loss is then reduced to 

$$\mathcal{L}_{MMD} = \|\frac{1}{N} \sum_{s=1}^N x^s - \frac{1}{M} \sum_{t=1}^M y^t \|^2,$$
in which $N$ and $M$ are the numbers of snapshot representations and temporal representations, respectively.


In the loss (Eqn.~\ref{eq:final-all}), the prediction loss can be different for different types of target datasets. For outcome prediction on chest radiographs, we prefer a combination of MMD and Cross-entropy loss as shown in Eqn.~\ref{eq:final1}. 
\begin{equation}
  \mathcal{L}_{chest} = \lambda_{1}\mathcal{L}_{MMD} + \mathcal{L}_{CE}
  \label{eq:final1}
\end{equation}

On the other hand, the severity changes in knee osteoarthritis disease is a classic application of ordinal regression~\cite{ordinal}. We use a combination of MMD and Ordinal loss as shown in Eqn.~\ref{eq:final2}.
\begin{equation}
  \mathcal{L}_{knee} = \lambda_{2}\mathcal{L}_{MMD} + \mathcal{L}_{Ord}
  \label{eq:final2}
\end{equation}

Next, we explain how the temporal representation and snapshot representation are obtained before recalibration.
\subsection{Temporal progression learning}
\label{ssec:method1}
Our temporal representation is obtained based on the state-of-the-art TCN. We will first provide some details of these techniques. For our problem, to match with snapshot representations, we need to find an optimal representation from the temporal sequence. To this end, we introduce a hierarchical self-attention module. The attention at different levels will be aggregated to obtain a single optimal representation for recalibration.

\myparagraph{Background: TCN.} The goal of the TCN is to gather the spatial dependencies over long ranges through causal dilated convolutions. The causal nature ensures that the computation of output at an individual timepoint $t$ 
depends only on the present and the past timepoints. Here, we briefly explain the temporal modeling using a TCN module. 

Assume a sequence of temporal images $\{x_{1}, x_{2},...,x_{T}\}$ available for each patient in the pretraining stage. 
This sequence of images is fed to a ResNet-18 model pretrained on ImageNet~\cite{resnet,imagenet}.
A vector with 512 elements is obtained with the extracted features. Considering $T$ temporal images of a patient, a sequence of 512-element vectors $p=\{p_{1}, p_{2},...,p_{T}\}$ is computed, which is then used as input to the TCN. 

\cref{architecture} shows the typical structure of a TCN with 2 dilated causal convolution layers. In our architecture we have used 3 such convolution layers. The dilation factors used for the layers are $d = \{1, 2, 4\}$ and the kernel size chosen is $k = 3$. Dilation is equivalent to introducing a fixed step between every $d$ adjacent filter taps. For a TCN layer, the relevant history information is obtained from $(k - 1)d$ past timepoints. 
For an input feature sequence $p$, and a filter $f : \{0,..., k-1\} \rightarrow \mathfrak{R}$, the dilated convolution $D$ on an element $s$ is
defined as:
\begin{equation}
  D(s) = \sum_{j=0}^{k-1} f(j) \cdot p_{(s-d.j)} 
  \label{eq:tcn}
\end{equation}
where $s-d.j$ refers to the past direction. 

\myparagraph{Hierarchical attention.}
In addition to feature extraction, we also exploit the TCN structure to select the optimal representation of the entire image sequence. As shown in \cref{architecture}, a multi-head self attention block is inserted between every two convolutional layers of the TCN. The input features are transformed into query $f$ and key $g$ via $1\times1$ convolutions. The attention map $A$ is then obtained from $f$ and $g$ by
\begin{equation}
  {A = \beta(f^{T}g})
  \label{eq:attweight}
\end{equation}
where $\beta$ is the softmax activation function. These maps contain self-attention weights which essentially quantify the importance of one timepoint relative to another. Next, the weighted representations $A^{T}h$ are generated, where $h$ is another set of feature transformed using $1\times1$ convolution. Finally we add the input features $p$, to $A^{T}h$. 

\begin{equation}
  {o = p + A^{T}h}
  \label{eq:attoutput}
\end{equation}

Note this output is $512\times T$ dimensional, consisting of representations for each of the $T$ time slices. For the final calibration, we need to use the attention $A$ to generate a 512 dimensional representation. 

Recall $A$ is a $T\times T$ matrix which is normalized for each column. We sum up each row of $A$ to get a $T$ dimensional weight vector $\alpha$, measuring the contribution of each time slice to all other time slices. A higher weight implies the corresponding time slice is more important. We apply softmax to this weight vector and use the output to calculate a weighted sum of the representations at all time slices. This final representation is then used in the recalibration task performed in section \ref{ssec:method2}. Further details on obtaining the optimal representation are provided in Supplementary. The pretraining of sequential images using TCN is performed with categorical cross entropy loss.

\subsection{Snapshot representation learning}
\label{ssec:method3}
For learning representations from snapshot images, we adopt the architecture of Self-supervised image Transformer (SiT)~\cite{sit} without the rotation task. We employ two tokens - first, the image patch token to perform image reconstruction, and second, the contrastive token of SimCLR~\cite{SS1} for contrastive prediction task.

\noindent\textbf{Reconstruction task.} For the image reconstruction task, a transformer is trained to extract the encoded visual features. The motivation is to learn context-preserving representations from the snapshot scans. Individual image grids are intentionally corrupted by passing through various transformations like addition of random noise, blurring by filters, and random grid replacement from another image. The transformer aims to restore the original image from the corrupted image. The output tokens of the transformer are aggregated to reconstruct the input image. The $\ell1$-loss between the input and the reconstructed image is employed as shown in Eqn. \ref{eq:recloss}:
\begin{equation}
\label{eq:recloss}
\mathcal{L}_{\rm r}(\mathbf{P}) = \frac{1}{D}\sum_i^D ||I_i - {\rm T}_{\rm r}(\bar{I_i})||
\end{equation}

where, $||.||$ is the $\ell1$ norm, $I_i$ is the original image, $\bar{I_i}$ is the corrupted image, ${\rm T}_{\rm r}(.)$ returns the reconstructed image and $D$ denotes the batch size.  $\textbf{P}$ are the parameters of the transformer to be learned during training.

\noindent\textbf{Contrastive learning task.} 
Positive pairs are constructed with the augmented versions of the same image. Augmentation techniques like cropping and horizontal flipping are used. Negative pairs denote samples coming from different inputs. The network is trained to minimize the distance between a positive pair and maximize the distance between a negative pair. This is achieved through a contrastive loss function, $\mathcal{L}_{\rm c}$, with cosine similarity as the similarity measure.
\begin{equation}
\label{eq:contrastive}
\mathcal{L}_{\rm c}(\mathbf{P}) = -\frac{1}{D}\sum_{i=1}^D \log \frac{{\rm e}^{{\rm sim}({\rm T}_{\rm c}(I_i), ~{\rm T}_{\rm c}({\tilde I_j}))/\tau}}{\sum_{j=1, j\neq i}^{2D} {\rm e}^{{\rm sim}({\rm T}_{\rm c}(I_i),~{\rm T}_{\rm c}(I_j))/\tau}}
\end{equation}
where ${\rm T}_{\rm c}(.)$ denotes the image embedding coming from the contrastive head, $\rm sim(.,.)$ is the dot product of the $\ell_2$ normalised inputs, which is the cosine similarity, and $\tau$ denotes a constant temperature parameter which we set to $0.5$. ${\tilde I_j}$ are $I_j$ are augmentations of the same image. The contrastive loss is defined as the arithmetic mean over all positive pairs in the batch of the cross entropy of their normalized similarities. For pretraining of snapshot images, we used a weighted combination of Reconstruction loss and Contrastive loss. It is given by:

\begin{equation}
  \mathcal{L}_{pre} = \lambda_{p1}\mathcal{L}_{r} +  \lambda_{p2}\mathcal{L}_{c}
  \label{eq:preloss}
\end{equation}
The pretrained representation will be used as the input for the recalibration network (Section \ref{ssec:method2}).

\begin{table*}[!b]
\begin{center}

\resizebox{0.8\textwidth}{!}{%
\begin{tabular}{*{9}{c}}
\hline
Name & \multicolumn{4}{c}{Ventilation} & \multicolumn{2}{c}{Mortality}  \\
\hline
Method & AUC($\uparrow$) & Sensitivity($\uparrow$) & Specificity($\uparrow$) & F1 score($\uparrow$) & AUC($\uparrow$) & Sensitivity($\uparrow$) & Specificity($\uparrow$) & F1 score($\uparrow$) \\
\hline\hline
COVID-Net \cite{comp1} & 0.73 & 0.64 & 0.69 & 0.70 & 0.76 & 0.63 & 0.73 & 0.72  \\
Rahimzadeh et. al \cite{comp2} & 0.75 & 0.62 & 0.74 & 0.68 & 0.77 & 0.58 & 0.76 & 0.74  \\
Oh et. al \cite{comp3} & 0.75 & 0.69 & 0.75 & 0.75 & 0.78 & 0.61 & 0.75 & 0.72  \\
COVIDiagnosis-Net \cite{comp4} & 0.71 & 0.60 & 0.71 & 0.68 & 0.78 & 0.56 & 0.74 & 0.65  \\
DarkCovidNet \cite{comp5} & 0.77 & 0.71 & 0.76 & 0.74 & 0.78 & 0.65 & 0.71 & 0.76  \\
\hline
CNN + LSTM \cite{comp6} & 0.74 & 0.63 & 0.66 & 0.63 & 0.76 & 0.62 & 0.69 & 0.69  \\
Azizi et. al \cite{tl2} & 0.79 & 0.74 & 0.79 & 0.76 & 0.80 & \textbf{0.74} & 0.77 & 0.78   \\
Li et. al \cite{comp8} & 0.79 & 0.71 & 0.75 & 0.74 & 0.81 & 0.68 & 0.76 & 0.80   \\
\hline
\textbf{Ours} & \textbf{0.88} & \textbf{0.76} & \textbf{0.80} & \textbf{0.79} & \textbf{0.87} & \textbf{0.74} & \textbf{0.78} & \textbf{0.84}  \\
\hline
\end{tabular}}
\end{center}
\vspace{-4mm}
\caption{COVID-19 outcome prediction results on SnapCXR dataset}
\label{tab:covidresults}
\end{table*}

\begin{table*}[t]
\begin{center}
\scalebox{0.8}{
\begin{tabular}{*{8}{c}}
\hline
Method & $\mu$F1($\uparrow$) & BA($\uparrow$) & AUC($\uparrow$) & Kappa($\uparrow$) & AUC(1,2)($\uparrow$) & AUC(2,3)($\uparrow$)  \\
\hline\hline
MobileNetV2 \cite{mobile} & 0.5104 & 0.3532 & 0.7822 & 0.2554 & 0.6208 & 0.6191  \\
CNN + Ordinal loss \cite{oc1} & 0.6865 & 0.6638 & 0.8950 & 0.5557 & 0.7298 & 0.8576 \\
SE block \cite{seblock} & 0.7336 & 0.7237 & 0.9237 & 0.6237 & 0.7866 & 0.9265 \\
Ensemble \cite{oc3} & 0.7405 & 0.7342 & 0.9112 &0.6327 & 0.7896 & 0.9360 \\ 
DeepKnee \cite{oc2} & 0.3956 & 0.5078 & 0.7456 &0.2287 &0.5931 &0.7398  \\
\hline
Ours w/o temporal (TCN+MMD) & 0.7844 & 0.7850 & 0.9520 & 0.6492  & 0.8428 & 0.9165\\
Ours w/o Attention & 0.8126 & 0.7607 & 0.9581 & 0.6842 & 0.8954 & 0.9126\\
Ours w/o Hierarchical Att & \textbf{0.8327} & 0.7883 & 0.9727 & 0.7065 & 0.8820 & 0.9284\\
\hline
\textbf{Ours} & 0.8265 & \textbf{0.8216} & \textbf{0.9773} & \textbf{0.7357} & \textbf{0.9167} & \textbf{0.9382} \\
\hline
\end{tabular}}
\end{center}
\vspace{-5mm}
\caption{Osteoarthritis severity prediction results and ablation studies on OA Kaggle dataset}
\label{tab:oairesults}
\end{table*}
\begin{table*}
\begin{center}
\resizebox{0.8\textwidth}{!}{%
\begin{tabular}{*{9}{c}}
\hline
Name & \multicolumn{4}{c}{Ventilation} & \multicolumn{2}{c}{Mortality}  \\
\hline
Method & AUC & Sensitivity & Specificity & F1 score & AUC & Sensitivity & Specificity & F1 score \\
\hline\hline
Ours w/o temporal (TCN+MMD) & 0.78 & 0.71 & 0.78 & 0.72 & 0.79 & \textbf{0.76} & 0.69 & 0.73  \\
Ours w/o Attention & 0.82 & 0.69 & 0.75 & 0.76 & 0.82 & 0.71 & 0.74 & 0.75  \\
Ours w/o Hierarchical Att & 0.86 & 0.75 & \textbf{0.84} & 0.74 & 0.85 & 0.75 & 0.71 & 0.80  \\
\textbf{Ours} & \textbf{0.88} & \textbf{0.76} & 0.80 &\textbf{0.79} & \textbf{0.87} & 0.74 & \textbf{0.78} & \textbf{0.84}  \\
\hline
\end{tabular}}
\end{center}
\vspace{-5mm}
\caption{Ablation study results for COVID-19 outcome prediction on SnapCXR dataset}
\label{tab:covidattablation}
\end{table*}
\begin{table}
\begin{center}
\resizebox{6cm}{!}{%
\begin{tabular}{*{7}{c}}
\hline
Name & \multicolumn{2}{c}{Ventilation} & \multicolumn{2}{c}{Mortality} & \multicolumn{2}{c}{OAI} \\
\hline
Method & AUC & F1 & AUC & F1 & AUC & $\mu$F1 \\
\hline\hline
KL & 0.81 & 0.67 & 0.82 & 0.77 & 0.92 & 0.76   \\
Bhattacharya & 0.77 & 0.65 & 0.75 & 0.71 & 0.89 & 0.79  \\
CORAL & 0.86 & 0.77 & 0.83 & 0.79 & 0.95 & \textbf{0.83} \\
\textbf{Ours (MMD)} & \textbf{0.88} & \textbf{0.79} & \textbf{0.87} & \textbf{0.84} & \textbf{0.97} & 0.82 \\
\hline
\end{tabular}}
\end{center}
\vspace{-5mm}
\caption{Ablation study results for different losses}
\label{tab:lossablation}
\end{table}

\section{Experiment design and results}
\label{sec:data}
To validate our proposed method, we perform experiments on two types of radiograph images. Chest radiographs (CXRs) are analyzed in order to predict clinical outcomes for COVID-19 patients and knee radiographs are studied to predict osteoarthritis (OA) severity.

The ability to predict clinical outcomes in COVID-19 can have significant implications on physician decision-making regarding medical resource allocation and treatment administration. This is particularly true in low-resource or surging case settings where triage must be performed. Similarly, grading of knee is clinically significant in monitoring a patient's disease progression and determining appropriate treatment measures. Furthermore, accurate grading of OA is critical for epidemiological surveys of disease prevalence. However, current scoring systems for OA rely heavily on physician interpretation of medical images which has been reported to have a high rate of inter-observer variability \cite{kohn2016classifications}. This might be overcome by automated approaches.

Multiple datasets are employed for model training and validation for the two applications listed above. While most datasets have only single-timepoint images, very few contain temporal data, described in detail below. 

\noindent\textbf{COVID-19 radiograph dataset.} For pretraining of the vision transformer, we used 28,433 images jointly from two Kaggle sources - 21,165 scans from a COVID-19-radiography-database \cite{data1} and 7268 from the SIIM-COVID19-detection challenge \cite{data2}. The temporal dataset, CovidProg, used for training TCN, included 942 images from 150 COVID patients. Evaluation was performed on the SnapCXR dataset, which had 631 snapshot images selected from TCIA \cite{data3} containing clinical outcome information.

\noindent\textbf{Knee osteoarthritis dataset.} We utilized two publicly available knee radiograph datasets, OAI \cite{kneedata1} and the Kaggle knee osteoarthritis dataset \cite{kneedata2}. OAI is a longitudinal study of 4,796 participants examined with radiographs and MRI during 9 follow-up examinations (0 to 96 months). Only the radiographs are studied in this work. Each radiograph also has an associated Kellgren-Lawrence (KL) score provided by a physician interpreter. There are 5 KL grades ranging from 0 to 4 that is a measure of the OA severity in knee joints.
For pretraining the transformer we jointly used 17,230 images from 4,350 OAI studies, and 5,778 images from the training folder of the Kaggle dataset. The remaining 426 OAI cases comprising 2,474 images were used as temporal data to train the TCN. Evaluation was done on a target dataset of 2,482 images utilizing both the validation and test folders in the Kaggle data. 

\subsection{Implementation details}
\noindent\textbf{Environment.} Our framework is built in Pytorch \cite{pytorch} and trained on an Nvidia GTX 2080Ti GPU. The self-supervised snapshot model is trained using the Adam optimizer. The learning rate and batch size are 0.0005 and 72, respectively. We used $\lambda_{p1}=\lambda_{p2}=1$. For optimization of the TCN model, we used the following parameters: learning rate = 0.001, momentum = 0.9, and optimizer=SGD. Pre-trained ResNet-18 features were used as input; the training converged within 20 epochs.

\noindent\textbf{Preprocessing.} For the COVID-19 chest radiograph datasets, lung region segmentation was first performed using a Residual U-Net model \cite{residual} so that features only from lung fields would be analyzed. All CXRs were aligned to the same intensity range through an average histogram matching method. For temporal data, each CXR image was divided into six grids. 
This was accomplished by dividing each lung into 3 equal zones. 
Corresponding grids of temporal images were used for separate timepoints, providing a grid-level registration across timepoints. Consequently, we obtained 6 training sequences for each patient in the temporal module. ResNet18 features were extracted from these grids and fed to TCN.

For the OAI dataset, we filtered out missing KL scoring labels and derived a training set of 19,704 knees. We utilized the 
BoneFinder tool \cite{BoneFinder} to localize the knee joint landmarks. Using these landmarks, the region of interest was cropped out for both knees. Following \cite{oc2}, histogram clipping and global contrast normalization were applied to each localized knee joint image. Finally, we rescaled all images to $310 \times310$ pixels using bilinear interpolation. The Kaggle OA dataset was used as provided. To feed the temporal data into TCN, each knee image was divided into two parts longitudinally and resized to dimension $128\times128$. Thus we obtained 4 training sequences for each patient in the temporal module. 

\noindent\textbf{Evaluation metrics.}
We employ AUC, sensitivity, specificity, F1 score to evaluate COVID-19 outcome prediction. t-distributed stochastic neighbor embedding (t-SNE) plots are computed for ventilation requirement prediction from features obtained after global average pooling. For knee severity multi-label classification, we used $\mu${F1}, balanced accuracy, AUC (one vs all), and Cohen's Kappa score as evaluation metrics. Images with the intermediate grades $1$, $2$ and $3$ are more difficult to differentiate from one another. Hence we also calculate AUC (one vs one) between grades 1,2 and 2,3 to measure their classification performance.

\subsection{Results}
\label{sec:result}
\begin{figure*}
  \begin{minipage}[t]{0.7\linewidth}
    \centering
    \includegraphics[scale=0.4]{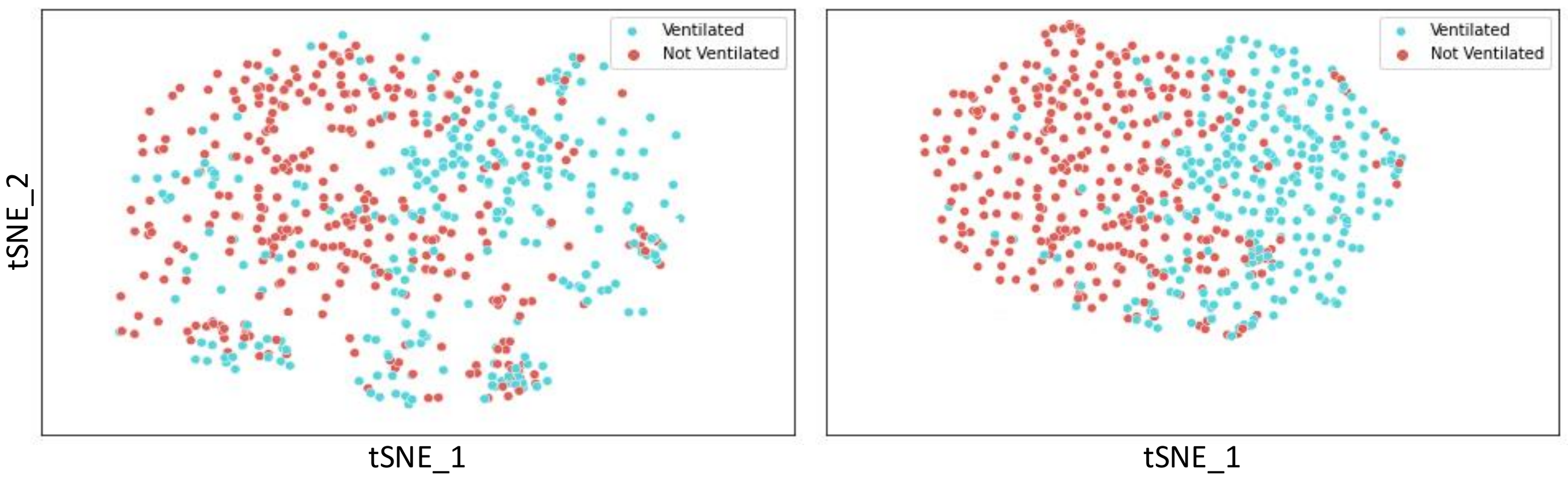}
    \caption{Comparison between t-SNE plots before and after using temporal modeling}
    \label{tempchange}
  \end{minipage}%
  \begin{minipage}[t]{0.3\linewidth}
    \centering
    \includegraphics[scale=0.33]{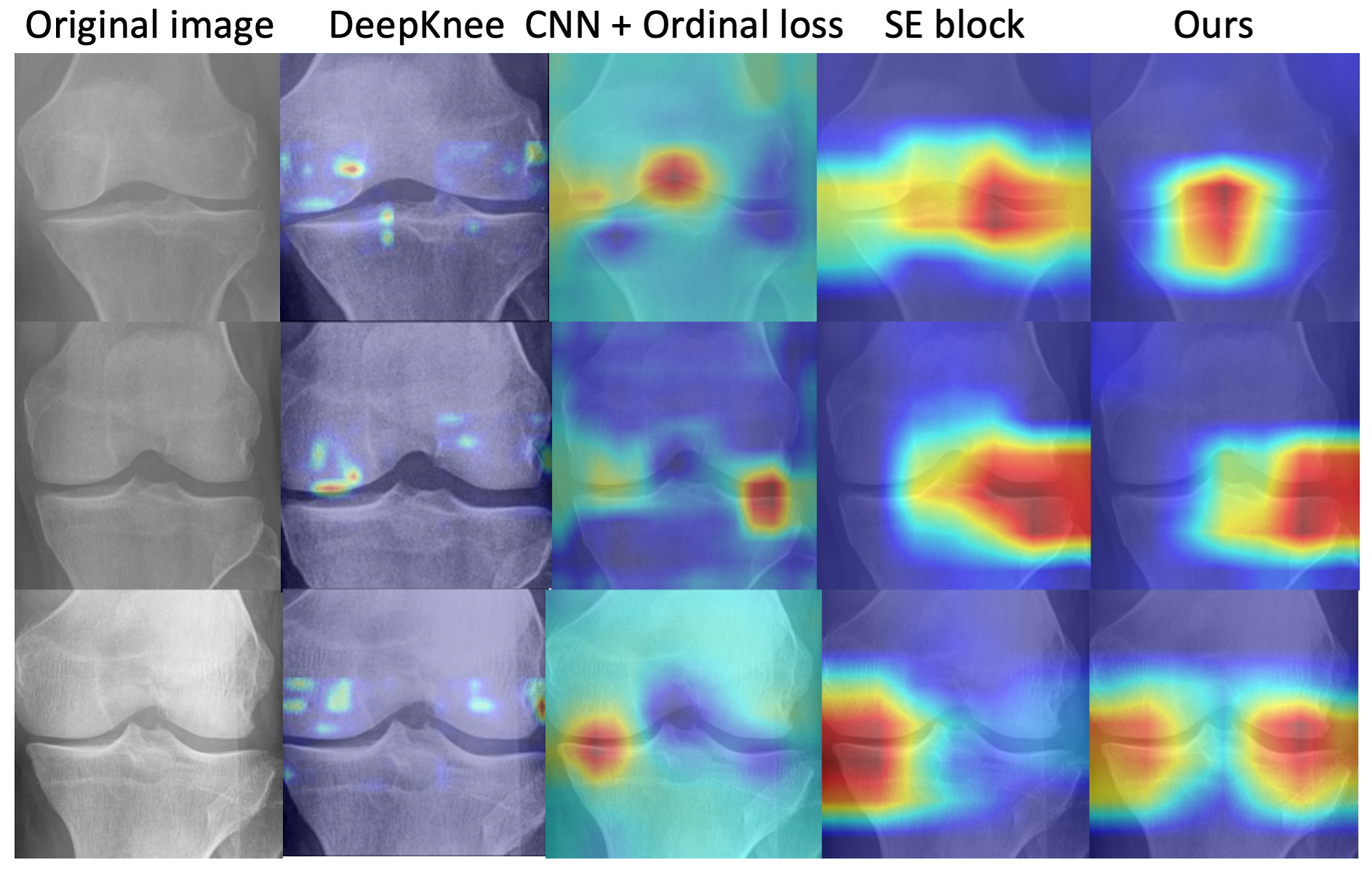}
    \caption{Qualitative comparisons \\of knee CAMs depicting OA severity}
    \label{gradcam}
  \end{minipage}
\end{figure*}

\begin{figure*}
  \begin{minipage}[t]{0.7\linewidth}
    \centering
    \includegraphics[scale=0.47]{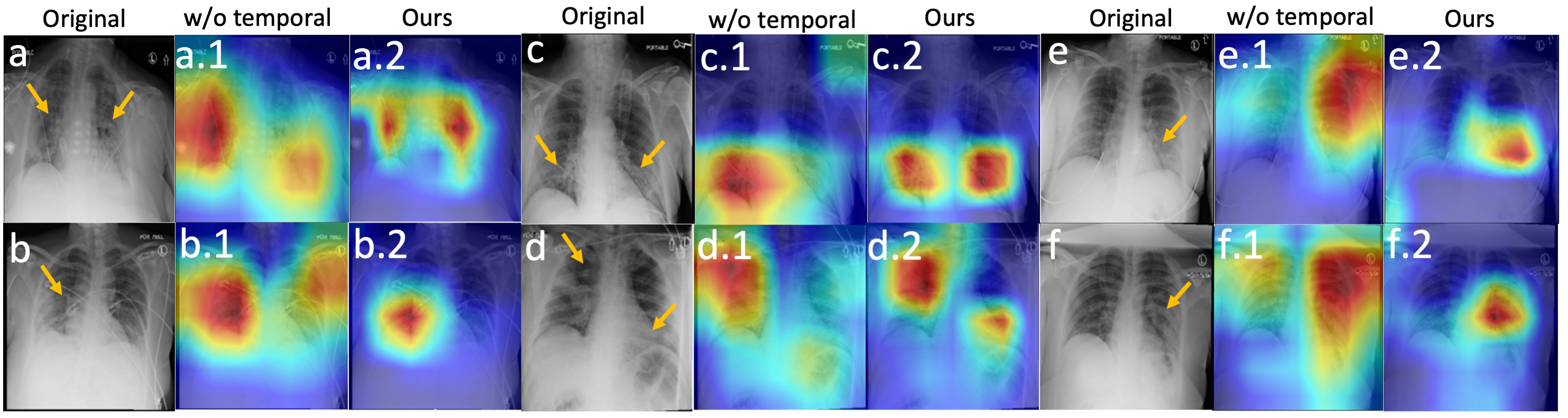}
    \caption{Qualitative comparison of chest CAMs before and after temporal modeling. a-f show CXRs with pathological lung infiltrates indicated by orange arrows a.1 - f.1 show generated CAMs before temporal modeling and a.2-f.2, after.}
    \label{cxrgrad}
  \end{minipage}%
   \hspace{0.4cm}
  \begin{minipage}[t]{0.3\linewidth}
    \includegraphics[scale=0.33]{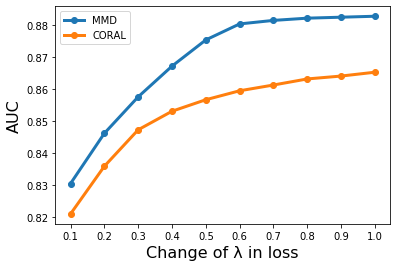}
    \caption{Optimization curve of feature distribution losses}
    \label{losscurve}
  \end{minipage}
\end{figure*}

\noindent\textbf{COVID-19 outcome prediction. Quantitative results:} 
To evaluate the proposed model in predicting COVID-19 induced mortality and mechanical ventilation, we compare it with seven methods including: COVID-Net \cite{comp1}, Rahimzadeh et. al \cite{comp2}, Oh et. al \cite{comp3}, COVIDiagnosis-Net \cite{comp4}, DarkCovidNet \cite{comp5}, CNN + LSTM \cite{comp6}, Azizi et. al \cite{tl2}, and Li et. al \cite{comp8}. Table \ref{tab:covidresults} shows that our method outperforms each of these methods on the SnapCXR dataset. COVID-Net \cite{comp1} uses a lightweight residual projection-expansion-projection-extension (PEPX) design pattern with convolutions at each stage. Rahimzadeh et. al \cite{comp2}, Oh et. al \cite{comp3}, and DarkCovidNet \cite{comp5} use various pretrained Deep CNNs. However, these fully supervised methods fail with limited training data, and share similar predictive performance. COVIDiagnosis-Net \cite{comp4} tuned a SqueezeNet architecture for COVID-19 diagnosis. Though they utilized an augmentation strategy, their lightweight backbone underperformed other state-of-art approaches. More recent methods, including Azizi et. al \cite{tl2} and Li et. al \cite{comp8}, exploit self-supervised learning strategies resulting in a boost in prediction performance. Our pipeline uses a self-supervised vision transformer as a feature extractor but also uniquely learns features of temporal progression. This approach outperforms other methods, achieving a $\sim11.5\%$ increase in AUC over state-of-the-art benchmarks for both ventilation and mortality prediction tasks.\textbf{ Qualitative results:} Fig. \ref{tempchange} demonstrates that utilizing temporal representations in our architecture results in better defined clusters between the ventilated and non-ventilated classes on the t-SNE plot. The inter-feature spatial distance also decreases leading to a more compact visualization. Class activation maps (CAMs) were generated before and after the inclusion of our temporal approach and are shown in Fig. \ref{cxrgrad}. \ref{cxrgrad}.b.2 and \ref{cxrgrad}.e.2 each demonstrate that our model more precisely localizes attention to pathological infiltrates when compared with baseline CAMs shown in \ref{cxrgrad}.b.1 and \ref{cxrgrad}.e.1. \ref{cxrgrad}.c.2 demonstrates improved localization of model attention to bilateral infiltrates rather than the large unilateral attention shown in \ref{cxrgrad}.c.1.

\noindent\textbf{OA severity prediction. Quantitative results:}
We compare our method against MobileNetv2 \cite{mobile}, CNN + Ordinal loss \cite{oc1}, Squeeze-Excitation block (SE block) \cite{seblock}, DeepKnee \cite{oc2} and, Ensemble \cite{oc3} in Table \ref{tab:oairesults}. MobileNetv2 \cite{mobile} incorporates bottleneck depth-separable convolution with residuals. 
CNN + Ordinal loss \cite{oc1} adds an ordinal loss for grade classification. This loss significantly improves the performance motivating us to leverage it in our framework. SE blocks performed best out of all comparative models achieving a 0.92 AUC.  Tiulpin et. al \cite{oc2} fused predictions from multiple Siamese deep models, but did not achieve good performance results. Our model is the first to leverage the temporal scans present in this OAI dataset to improve severity grading. Learning the progression of OA vastly improved the classification results from single images. We achieved a $\sim5.8\%$ increase in AUC over the state-of-the-art benchmark \cite{seblock}. Our model also outperforms state-of-the-art approaches in the clinically difficult problem of discriminating between intermediate grades 1 vs 2 and 2 vs 3 (given by AUC(1,2) and AUC(2,3)). We also reported a Cohen’s Kappa value of 0.73 for our method which was the best among all approaches, demonstrating our model's higher agreement with ground truth KL scores. \textbf{Qualitative results:} CAMs generated from the last layer of each compared model are illustrated in Fig. \ref{gradcam}. Both our model and the SE block baseline \cite{seblock} show superior localization of attention to pathological osteophyte development and joint-space narrowing when compared with other baselines. The Supplementary section contains further qualitative analyses. 

\subsection{Ablation Study}
\label{sec:ablation}
\noindent\textbf{Effectiveness of Attention:} Several experiments are performed on the SnapCXR and OA Kaggle datasets
to verify the benefit of each component of our proposed framework. We first remove the temporal module from the framework in \cref{architecture} and perform predictions from only snapshot images using the SiT \cite{sit}. This network is taken as the baseline feature extractor without TCN and Attention. Compared with this baseline, our model yields a $12.8\%$ and $10.12\%$ AUC improvement in ventilation and mortality prediction, and a $\sim2.65\%$ improvement for OA severity prediction. We then average only the representations from the output of TCN without using attention, inhibiting the model from choosing an optimal representation from the temporal sequence. It can be seen in Table \ref{tab:covidattablation} that introducing the Global Attention module improves our AUCs to 0.86 and 0.85 for ventilation and mortality prediction on SnapCXR, and to 0.97 AUC for OA severity classification on the Kaggle dataset. Finally, our application of hierarchical self-attention performs slightly better than the global attention-weighted features. We attribute this improvement to the self-attention exploited after every layer, enabling better contextual understanding of the modality itself. 
\noindent\textbf{Effectiveness of MMD loss:}
Table \ref{tab:lossablation} presents ablation results for different loss functions. Using the MMD loss resulted in $8.6\%,6.09\%$ performance improvement (AUC) for SnapCXR prediction tasks and $5.4\%$ improvement in OA severity prediction in the Kaggle dataset. KL-divergence and Bhattacharya losses help in matching two posterior probability distributions and result in inferior performance. Because our distributions are derived from two different contexts (temporal vs snapshot) used for two different purposes (severity progression tracking and clinical outcome prediction), feature level losses like MMD and CORAL are more appropriate. Fig. \ref{losscurve} shows the AUC achieved at different stages by varying the weights of MMD and CORAL in the final loss for our method and ablation work, respectively. The MMD AUC curve achieves saturation earlier at $\lambda=0.5$, suggesting that even a small weight of MMD loss improved results.


\vspace{-1mm}
\section{Conclusion}
This paper presents a novel framework for the augmentation of snapshot-image--based pipelines by the integration of information from multi-image sequences. 
Unlike existing approaches trained only on snapshot images, our architecture learns disease progression representations which are used to re-calibrate snapshot features. When evaluated on chest and knee radiograph datasets, the proposed architecture outperforms state-of-the-art approaches. 
This work paves the way for the inclusion of temporal data as auxiliary information for single image-based training paradigms.

\vspace{-1mm}
\section{Acknowledgements}
Reported research was supported by the OVPR and IEDM seed grants at Stony Brook University, NIGMS T32GM008444, and NIH 75N92020D00021 (subcontract). The content is solely the responsibility of the authors and does not necessarily represent the official views of the National Institutes of Health.

{\small
\bibliographystyle{ieee_fullname}
\bibliography{egbib}
}

\section*{Supplementary Material}
In this supplementary material, we provide additional information to further understand our proposed approach. In section \ref{tcncalattn}, we provide an architectural overview of how to calculate a 512 dimension vector from last the layer of Temporal Convolutional Network (TCN). We provide dataset and preprocessing details in section \ref{data}. Finally in section \ref{oaexp}, t-SNE plots and additional class activation maps provide insights into OA severity prediction from knee radiographs.

\section{Architecture details - TCN output}
\label{tcncalattn}

\begin{figure}[ht]
  \centering
  \includegraphics[width=0.9\linewidth]{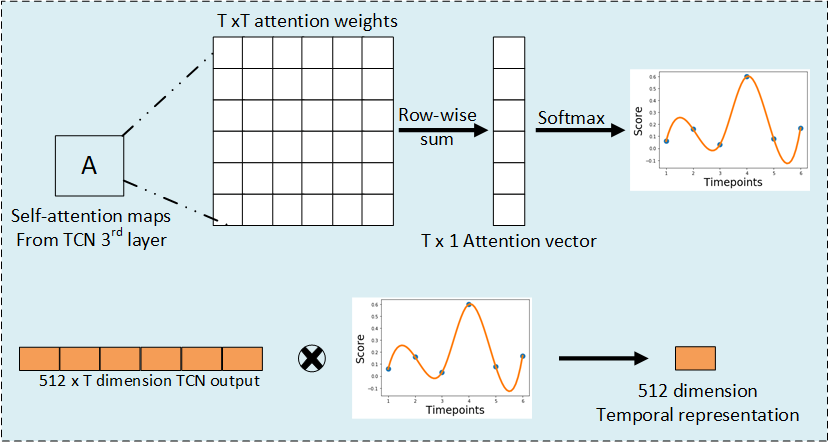}
   \caption{Derivation of temporal representation from last layer of TCN.}
  \label{tcncalc}
\end{figure}

A row-wise summation operation is applied on the self-attention weights obtained from the third and final self-attention block in our TCN architecture. This results in a `T' length attention vector, the softmax of which gives the attention scores. When these attention scores are multiplied with the output of TCN, they generate an optimal 512 dimension representation. The steps are illustrated in Supplementary Fig \ref{tcncalc} 

\section{Assets and preprocessing}
\label{data}

\noindent\textbf{Chest Radiograph dataset:}
For snapshot pretraining, we used 28433 chest radiographs (comprising multiple pulmonary diseases). CovidProg dataset, which contained 942 scans from 150 COVID-19 patients, comprised the temporal data. The duration between the CXR scans are variable (1-5 days). The number of timepoints per patient varies from 4 to 16. Out of the total 150 patients, 23 cases were obtained from Newark Beth Israel Medical Center, 77 from Stony Brook University Hospital, and 50 from University Hospitals Cleveland Medical Center.

\noindent\textbf{OA Radiograph dataset:}
For snapshot pretraining, we used 23008 images. 2474 knee scans from 426 patients comprised the temporal data. The images in the `train' folder of Kaggle \cite{kneedata2} were a fraction of the snapshot cohort used in pretraining the transformer. All the images in `validation' and `test' folder \cite{kneedata2} were jointly used in finetuning stage. 

The experiments were performed in a 5-fold cross validation setting in the finetuning stage where the pretrained transformer model was finetuned on 4 folds and tested on the remaining fold. Details about the data used for each stage can be found in Supplementary Table \ref{tab:datasplit}

\begin{table}[ht]
\begin{center}
\resizebox{8cm}{!}{%
\begin{tabular}{*{3}{c}}
\hline
Stage & COVID & OA \\
\hline
Snapshot & 21165\cite{data1} + 7268 \cite{data2} & 17230 \cite{kneedata1} + 5778 train folder \cite{kneedata2}  \\
Temporal & 942 (CovidProg) & 2474\cite{kneedata1}  \\
Finetune & 631 (Vent.), 531 (Mort.) \cite{data3} & 2482 val + test folder\cite{kneedata1}  \\
\hline
\end{tabular}}
\end{center}
\vspace{-5mm}
\caption{Data utilized in different stages}
\label{tab:datasplit}
\end{table}

\noindent\textbf{COVID-19 preprocessing:}
Lung region segmentation was first performed using a Residual UNet model \cite{residual}. All chest scans were aligned to the same intensity range through an average histogram matching method.


\noindent\textbf{OA preprocessing:} We utilized BoneFinder tool \cite{BoneFinder} to localize and crop the knee joint landmarks. Following \cite{oc2}, histogram clipping and global contrast normalization were
applied to each localized knee joint image.

Samples of COVID and OA images after pre-processing are shown in Supplementary Fig \ref{prechest} and Supplementary Fig \ref{bfoutput}, respectively.

\begin{figure*}[ht]
  \centering
  \includegraphics[width=0.9\linewidth]{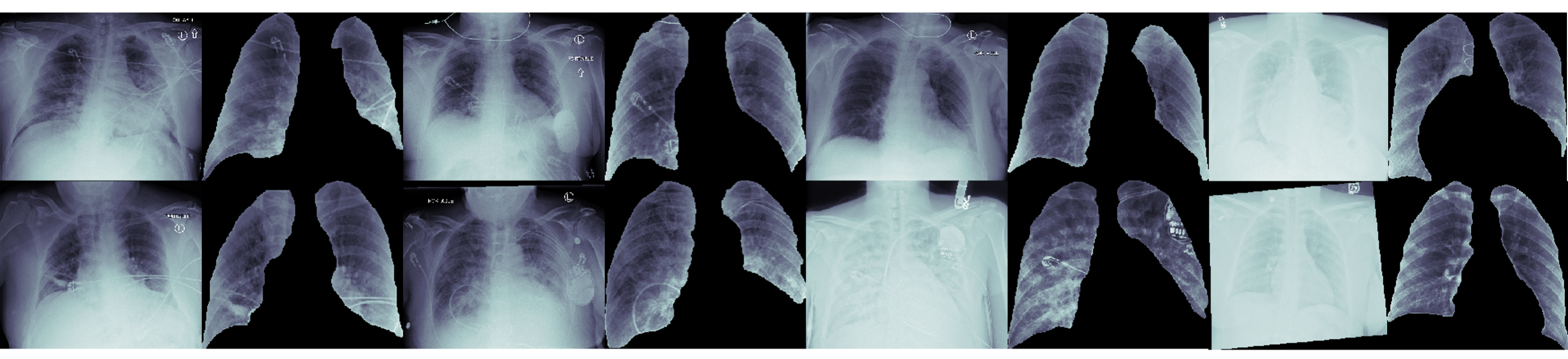}
    \caption{Preprocessed chest scans after applying average histogram matching and lung segmentation}
    \label{prechest}
\end{figure*}

\begin{figure}[h]
  \centering
  \includegraphics[width=0.6\linewidth]{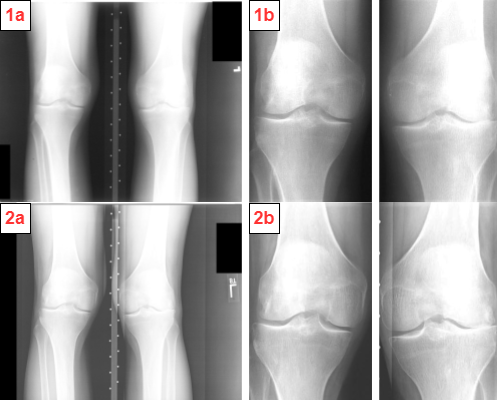}
    \caption{Preprocessed knee radiographs (1b, 2b) generated after joint localization \cite{BoneFinder} and global contrast normalization \cite{oc2} on original samples (1a, 2a)}
    \label{bfoutput}
\end{figure}

\begin{figure*}[h]
  \centering
  \includegraphics[width=0.85\linewidth]{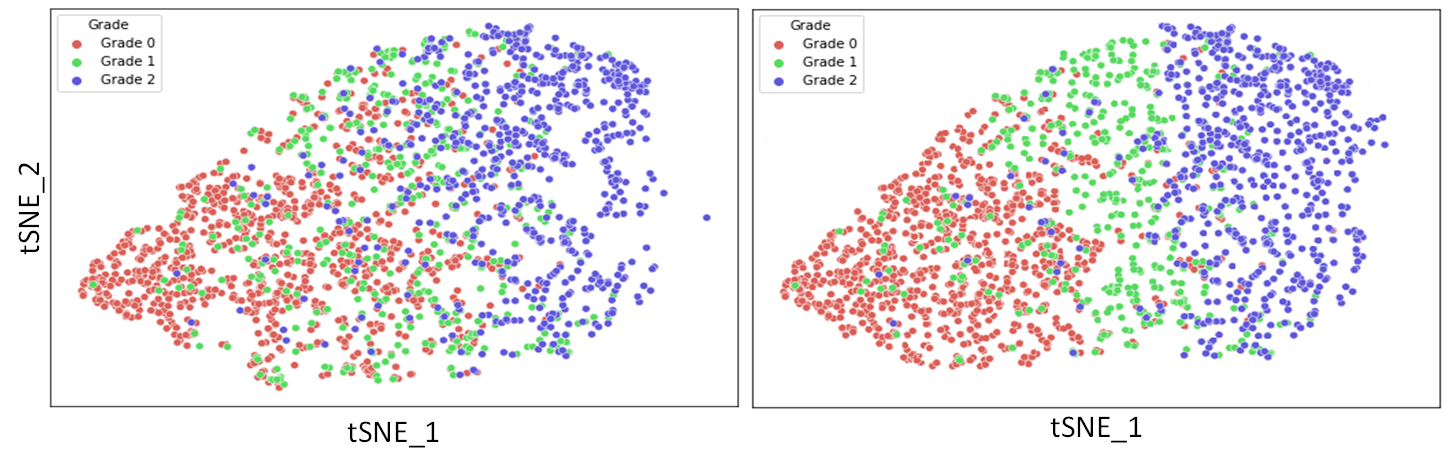}
    \caption{Comparison between t-SNE plots before and after using temporal modeling for severity grades (0,1,2)}
    \label{tempoai}
\end{figure*}

\section{Insights from OA severity prediction}
\label{oaexp}
Supplementary Fig \ref{tempoai}  demonstrates that  utilizing temporal  representations  in our  architecture  results  in better defined  clusters for the three severity grades $(0,1,2)$ on the t-SNE plot. It may be observed that intermediate grades, such as 1 vs 2, which are more difficult to predict (left) can benefit from the proposed temporal approach (right). Additional CAMs of OA affected knees are compared in Supplementary Fig \ref{kneecamextra}. Each row corresponds to knee radiographs from different severity grades, from 0 to 4. As may be observed, the attention maps from DeepKnee \cite{oc2} , CNN + Ordinal loss \cite{oc1} and SE block \cite{seblock} are very sparse and sometimes react to unnecessary areas (bone texture, joint centre). On the contrary, our method provides more focused attention on the osteophytes and joint narrowing - the two important indicators of osteoarthritis.  

\section{Recalibration using matching data}
\label{matchexp}
In the COVID-19 cohort, we included some matched data. 100 of 150 patients in the CovidProg temporal dataset also have their ventilation status known. We use the 100 patients, take the baseline scans (the first image) of their temporal sequences as matched snapshot images. 
We evaluated the distance between these matched temporal/snapshot data in the representation space through training. In Supplementary Fig \ref{distfeature}, Curve A (blue) shows the average distance ($d$) between the matched pair of snapshot and temporal representation. $d$ is reduced to only 0.10 after 40 epochs. For reference, we also show $d$ between any snapshot of positive ventilation status (S+) and any temporal sequence of positive ventilation status (Tm+). The result is Curve B (orange). Meanwhile, we also report $d$ between S+ and any temporal sequence with negative ventilation status (Tm-) as in Curve C (green). After 40 epochs, $d$ in A,B,C are 0.10, 0.67 and 4.26, respectively. C$>>$B$>$A shows that (1) the matched snapshots and temporal sequences are automatically aligned very well during training, thanks to the MMD loss; (2) generally a positive temporal sequence is aligned much closer to a positive snapshot than a negative snapshot, although not as close as the matched pairs.

\section{Longitudinal comparison}
We also compared our method with other temporal models, namely CNN+LSTM, CNN+biLSTM and CNN+biLSTM+Attention. It may be observed from Supplementary Table \ref{tab:longablation} that our approach outperforms all these longitudinal models. 

\begin{table}[H]
\begin{center}
\resizebox{6cm}{!}{%
\begin{tabular}{*{7}{c}}
\hline
Name & Ventilation \\
\hline
Method & AUC \\
\hline\hline
CNN+LSTM & 0.82   \\
CNN+biLSTM & 0.83  \\
CNN+biLSTM+Attention & 0.85  \\
\textbf{Ours} & \textbf{0.88} \\
\hline
\end{tabular}}
\end{center}
\vspace{-5mm}
\caption{Comparison with longitudinal methods}
\label{tab:longablation}
\end{table}

\section{Limitations}
In our temporal analysis, the images are not registered. Registration might result in learning better representations. We aim to address this in the future by using a spatial transformer network as a pre-processing stage before extracting temporal features. Also, due to lack of sufficient temporal data, we did not use transformer architectures to learn disease progression. This research direction can be pursued with the availability of more temporal cases in future.

\begin{figure}[H]
  \centering
  \includegraphics[width=0.8\linewidth]{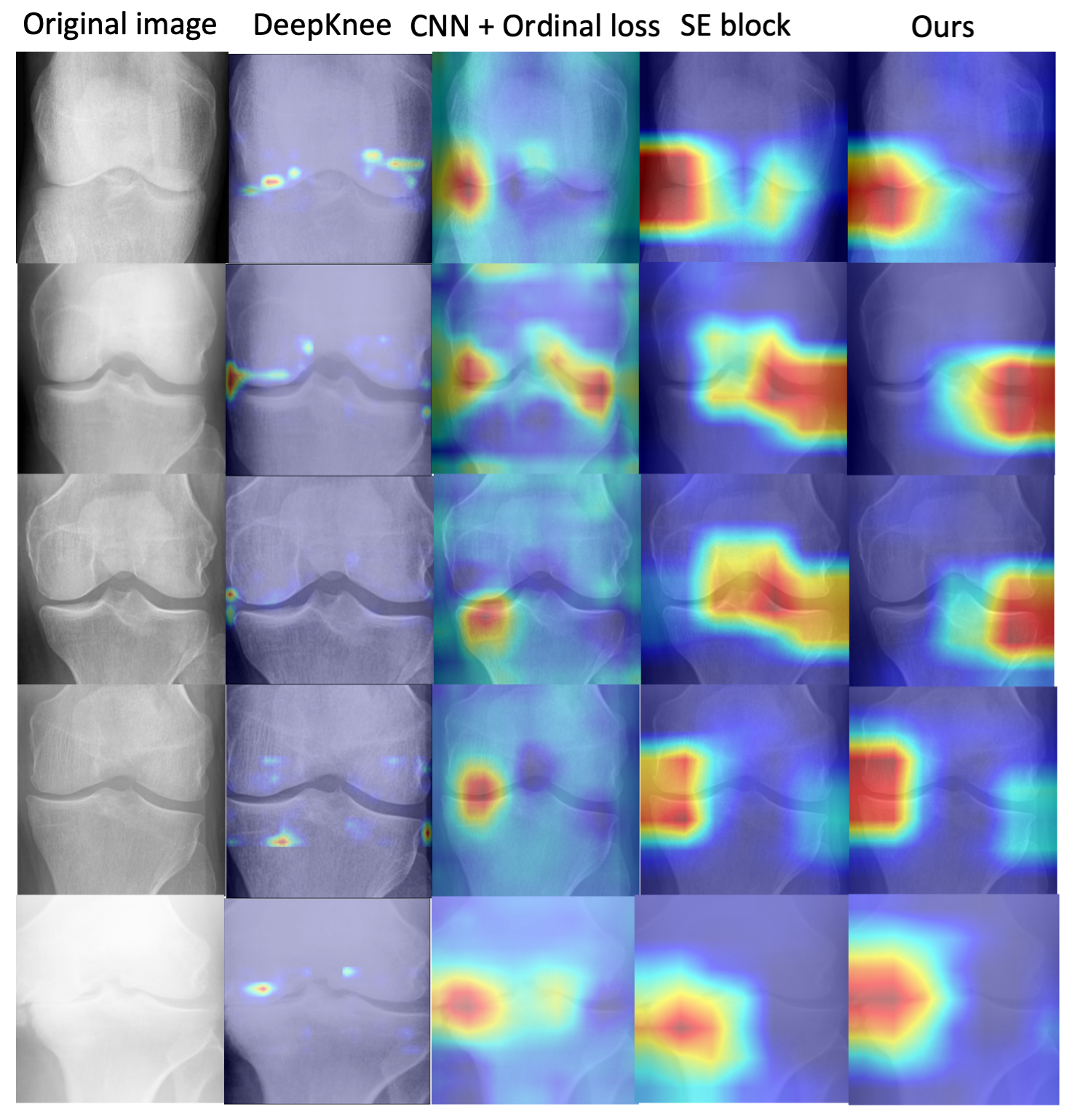}
    \caption{Qualitative comparisons of knee CAMs depicting OA severity grades 0 to 4 (top to bottom)}
    \label{kneecamextra}
\end{figure}

\begin{figure}[H]
  \centering
  \includegraphics[width=0.8\linewidth]{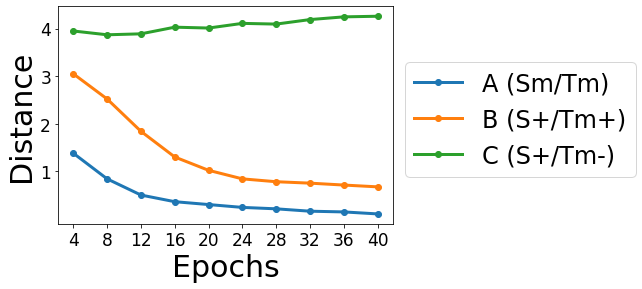}
  \caption{{Distance between feature means across training}}
  \label{distfeature}
\end{figure}



\end{document}